\documentclass[12pt, draftclsnofoot, onecolumn]{IEEEtran}
\usepackage{amsmath,cite,bm,amsfonts,amssymb,graphicx,color,mathtools,setspace,comment,multirow,xfrac}
\usepackage{epstopdf}
\onecolumn
\newcommand{\bH}{\mathbf{H}}

\newcommand{\bG}{\mathbf{G}}
\newcommand{\bg}{\mathbf{g}}
\newcommand{\bI}{\mathbf{I}}

\newcommand{\bE}{\mathbf{E}}
\newcommand{\be}{\mathbf{e}}
\newcommand{\bR}{\mathbf{R}}

\newcommand{\bu}{\mathbf{u}}
\newcommand{\bP}{\mathbf{P}}
\newcommand{\bX}{\mathbf{X}}
\newcommand{\bQ}{\mathbf{Q}}
\newcommand{\bv}{\mathbf{v}}

\newcommand{\bw}{\mathbf{w}}

\newcommand{\bx}{\mathbf{x}}
\newcommand{\bs}{\mathbf{s}}
\newcommand{\bq}{\mathbf{q}}

\newcommand{\raisecaption}{\vspace{-0.7cm}}
\ifCLASSINFOpdf
\else
\fi
\begin{document}
\doublespacing
\title{On the Convergence and Performance of MF Precoding in Distributed Massive MU-MIMO Systems}
\author{\IEEEauthorblockA{Peter J. Smith\IEEEauthorrefmark{1},
											Callum T. Neil\IEEEauthorrefmark{2},
											Mansoor Shafi\IEEEauthorrefmark{3},
											Pawel A. Dmochowski\IEEEauthorrefmark{2}
											}
\\
\IEEEauthorblockA{\IEEEauthorrefmark{1}
\small{Department of Electrical and Computer Engineering, University of Canterbury, Christchurch, New Zealand}} \\
\IEEEauthorblockA{\IEEEauthorrefmark{2}
\small{School of Engineering and Computer Science, Victoria University of Wellington, Wellington, New Zealand}} \\
\IEEEauthorblockA{\IEEEauthorrefmark{3}
\small{Spark New Zealand, Wellington, New Zealand}}
\IEEEauthorblockA{\small{email:~p.smith@elec.canterbury.ac.nz,~\{pawel.dmochowski,callum.neil\}@ecs.vuw.ac.nz,\\ mansoor.shafi@spark.co.nz}}}

\maketitle

\begin{abstract}
In this paper, we analyze both the rate of convergence and the performance of a matched-filter (MF) precoder in a massive multi-user (MU) multiple-input-multiple-output (MIMO) system, with the aim of determining the impact of distributing the transmit antennas into multiple clusters. We consider cases of transmit spatial correlation, unequal link gains and imperfect channel state information (CSI). Furthermore, we derive a MF signal-to-interference-plus-noise-ratio (SINR) limit as both the number of transmit antennas and the number of users tend to infinity. In our results, we show that both the rate of convergence and performance is strongly dependent on spatial correlation. In the presence of spatial correlation, distributing the antennas into multiple clusters renders significant gains over a co-located antenna array scenario. In uncorrelated scenarios, a co-located antenna cluster has a marginally better mean per-user SINR performance due to its superior single-user signal-to-noise-ratio (SNR) regime, i.e., when a user is close to the base station (BS), the links between the user and all transmit antennas becomes strong.
\end{abstract}
\section{Introduction}
It is well known that increasing the number of antennas at the base station (BS) can result in large increases in data rate, reliability, energy efficiency and reduced inter-user interference \cite{LARSSON}. Consequently, massive multiple-input-multiple-output (MIMO) is being investigated as an emerging technology \cite{LARSSON,RUSEK,MARZETTA,NGO,PITAROKOILIS,HOYDIS,MULLER,HUH,YANG,OZGUR,JOSE,NGO2,NGO3,CHOI,CHOI2,FERNANDES,TRUONG,GAO}, where the number of antennas is scaled up by many orders of magnitude relative to systems today. Performance benefits from such a large number of antennas include an improvement in radiated energy efficiency of 100 times \cite{LARSSON} relative to single-antenna, single-terminal systems. In \cite{NGO}, the authors demonstrate that using linear processing at the transmitter you can achieve a spectral efficiency improvement of up to two orders of magnitude while simultaneously improving energy efficiency by three orders of magnitude.
\par
The analysis of precoding techniques for massive MIMO has been the subject of a number of studies such as \cite{YANG,NGO3,HOYDIS,JOSE,NGO,TRUONG,RUSEK,GAO}. Conjugate beamforming (BF) (matched filter (MF)) and pseudo inverse BF (zero forcing (ZF)) precoding methods were considered in \cite{YANG}, comparing spectrum efficiency with radiated efficiency. In \cite{NGO3}, capacity expressions were derived for maximum ratio transmission (MRT) and ZF techniques, including scenarios with channel estimation imperfections. Channel state information (CSI) imperfections were also considered in \cite{NGO2,NGO3,CHOI,CHOI2,FERNANDES,TRUONG}. In \cite{NGO3}, the authors propose a BF training scheme to acquire CSI by means of short pilot sequences while maintaining a low channel estimation overhead. The effects of channel aging on CSI in massive MIMO systems were looked at in \cite{TRUONG}, where the authors derive achievable rates for uplink (UL) and downlink (DL) when channel aging effects, modeled using a first order autoregressive process, were considered for MF precoders. The paper compares achievable rates of perfect CSI, aged CSI and predicted CSI.
\par
As the channel matrix dimension becomes large, the analysis of massive MIMO systems is aided by random matrix theory asymptotics. The effect of increasing array size has been the subject of a few studies, e.g., \cite{RUSEK,SMITH2}. In \cite{RUSEK}, the authors conclude that effects of random matrix theory are observable even for arrays of 10 antennas, although the desirable properties of an ``infinite'' number of antennas are more prominent at 100 antennas and above. The convergence of random matrix theory asymptotics is shown via simulation in \cite{SMITH2}, as the number of BS antennas is increased. \cite{SMITH2} concludes that the number of antennas required to achieve equal singular values is well over $10^{4}$. Practical simulations, in \cite{GAO}, provide measurements in residential areas with 128 BS antennas, which shows that orthogonality improves for an increasing number of antennas, but for a system with two single-antenna users, little improvement beyond a 20 antenna element array is seen.
\par
In adding more antennas to a fixed array size, distances between adjacent elements are reduced. In a massive MIMO system, the effects of inter-element spatial correlation are increased dramatically \cite{ZHANG,ZHANG2,MATTHAIOU,YIN2}, due to the significant reduction in antenna spacing. However, this could be partially mitigated by dividing the antennas into multiple clusters whereby antenna spacings per cluster increase provided the overall form factor remains the same as the co-located BS case. The primary aim of this paper is to analyse the performance of a massive MIMO system by distributing the antenna elements into multiple clusters. Specifically, our motivation is analysing per-user MF signal-to-interference-plus-noise-ratio (SINR) as the number of antenna elements becomes large.
\par
In this paper we analyze the MF precoding technique in a \textit{distributed} BS scenario. Our contributions can be summarized as follows: 
\begin{itemize}
	\item We provide a system model for a massive multi-user (MU)-MIMO system which accounts for: distributed transmit antennas, unequal link-gains between users and antenna clusters, CSI imperfections, and transmit spatial correlation, from which we analyze the MF precoding technique and derive analytical expressions for expected per-user MF SINR.
	\item We analyse the impact of different numbers of antenna clusters on spatial correlation and expected per-user MF SINR.
	\item We analytically derive a limiting expected per-user MF SINR and show via simulation the convergence of the instantaneous per-user MF SINR to this limit.
\end{itemize}
\par
The remainder of this paper is organized as follows. First, in Section \ref{SystemModel}, we describe the system model and assumptions. In Section \ref{SINRAnalysis}, we derive the expected per-user MF SINR and the limit as the number of BS antennas and the number of single-antenna users increase without bound, at a constant ratio. Then, in Section \ref{NumericalResults}, we present numerical simulations and show the impact of distributing the antennas into multiple clusters. The majority of the mathematical derivations are included in the Appendices.
%
\section{System Model}
\label{SystemModel}


\subsection{Precoding}
\label{Precoding}
We consider a massive MIMO DL system with a total of $M$ transmit antennas divided equally among $N$ BSs (antenna clusters), jointly serving a total of $K$ single-antenna users. At each BS the $\frac{M}{N}$ antennas are assumed to be arranged as $\frac{M}{2N}$ pairs of cross-polarized (x-pol) antennas. We assume time division duplex (TDD) operation with UL pilots enabling the transmitter to estimate the DL channel. On the DL, the $K$ single antenna terminals collectively receive the $K\times 1$ vector
\begin{equation} \label{x}
	\bx = \sqrt{\rho _{\textrm{f}}}\bG^{\textrm{T}}\bs + \bw,
\end{equation}
where $\rho _{\textrm{f}}$ is the transmit signal-to-noise-ratio (SNR),  $\bs$ is an $M \times 1$ precoded data vector and $\bw$ is a $K\times 1$ noise vector with independent and identically distributed (i.i.d.) $\mathcal{CN}(0,1)$ entries. The transmit power is normalized, $\mathbb{E}\left[ \| \bs\| ^{2}\right] = 1$, i.e., each antenna transmits at a power of $\frac{\rho _{\textrm{f}}}{M}$. The $M\times K$ channel matrix, $\bG$, is given by
\begin{equation} \label{G}
	\bG = \begin{bmatrix}	
		\beta _{1,1}^{\frac{1}{2}}\bR_{\textrm{t}}^{\frac{1}{2}}\bH_{1,1}	&	\ldots 	&	\beta _{1,K}^{\frac{1}{2}}\bR_{\textrm{t}}^{\frac{1}{2}}\bH_{1,K}	\\
       		\vdots    			  		 			& 	\ddots 	& 	\vdots    			   		 			\\
       		\beta _{N,1}^{\frac{1}{2}}\bR_{\textrm{t}}^{\frac{1}{2}}\bH_{N,1} & 	\ldots 	& 	\beta _{N,K}^{\frac{1}{2}}\bR_{\textrm{t}}^{\frac{1}{2}}\bH_{N,K}		
     	\end{bmatrix},
\end{equation}
where $\bH_{n,k}\in \mathbb{C}^{\frac{M}{N}\times 1}$, with i.i.d. $\mathcal{CN}(0,1)$ entries, is the channel vector between the $k$th user and the $n$th BS, corresponding to small-scale Rayleigh fading. $\beta _{n,k}$ is the link gain coefficient, modeling large-scale effects for user $k$ from BS $n$, while $\bR_{\textrm{t}}$ is the spatial correlation matrix at each antenna cluster, assumed equal for all BSs.
\par
In this paper, we consider the convergence scenario where $K,M\rightarrow \infty $ with a fixed ratio of $\alpha = \frac{M}{K}$, where cases of $N=1,2$ and $N>2$ are examined. Note that although we consider finite $N$, the analysis can also be extended to the case where $N\to \infty $.
\subsection{Link Gain Model}
\label{LinkGainModel}
With distributed users and distributed antenna clusters the link gains, $\beta _{n,k}$, in a real system will all be different due to variations in path-loss and shadowing. In this paper, we have two areas of interest: massive MIMO performance and convergence. Hence, we model the link gains in two different ways.
\subsubsection{Statistical Link Gain Model}
\label{StatisticalLinkGainModel}
Here, we adopt the classical model where users are dropped at random locations in a circular coverage area served by the antenna clusters. The link gains are then generated assuming i.i.d. log-normal shadow fading and distance based path-loss. Since each drop generates substantially different link gains, this model is not ideal for investigating convergence as the link gain variations may confound the limiting effects. However, the model is useful for simple generation of arbitrary system sizes and can be used to investigate massive MIMO performance for a widely accepted link gain model. Finally, the limiting results can be compared to the SINR of an individual drop to evaluate the accuracy of the limit as an approximation to a particular massive MIMO system.
\subsubsection{Limiting Link Gain Model}
\label{LimitingLinkGainModel}
\begin{figure}[ht]
\centering\includegraphics[trim=1cm 8.5cm 21cm 1cm,clip,width=0.7\columnwidth]{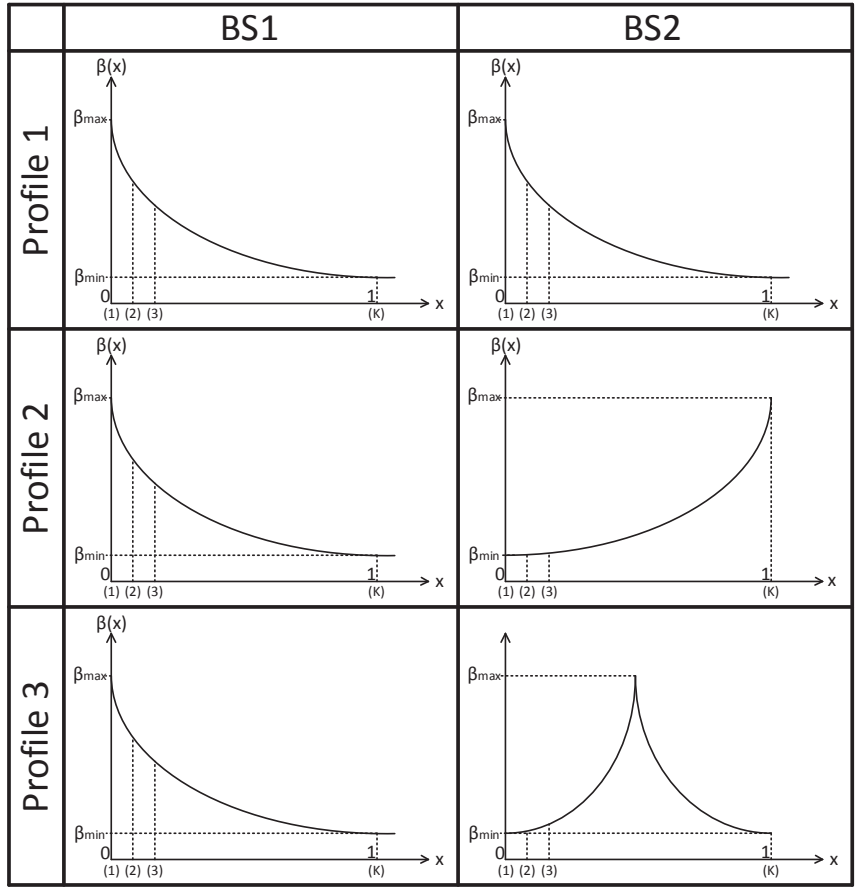}
\raisecaption\caption{Link gain profiles for a two BS scenario where $(1),(2),(3)\ldots (K)$ represent the users.}
\label{LinkGainProfiles}
\end{figure}

Here, we assume that the link gains between an antenna cluster and $K$ users are drawn from a limiting link gain profile defined by $\beta (x)$ for $0\leq x\leq 1$. For any finite number of users, $K$, the link gains are defined by $\beta ((2k-1)/2K)$ for $k=1,2,\ldots ,K$. For the first antenna cluster, we use the model, $\beta (x)=\beta _{\textrm{max}}(\beta _{\textrm{min}}/\beta _{\textrm{max}})^{x}$, where $\beta _{\textrm{min}}$ and $\beta _{\textrm{max}}$ are the minimum and maximum link gains respectively. This simple model also appears in \cite{SMITH2,BASNAYAKA} as a way of characterizing differing user link gains with a simple exponential profile and only two parameters. The resulting link gain profile is shown in Figure \ref{LinkGainProfiles} under BS1. For simplicity, we also assume that the second cluster (BS2) has the same link gain profile. However, it is unrealistic to assume that the same users have the same link gains at both BSs. Hence, we consider three scenarios, labeled Profile 1, 2 and 3 in Figure \ref{LinkGainProfiles}. In Profile 1, both BSs have the same profile to all $K$ users. In Profile 2 the profiles for BS1 and BS2 are reversed, so that a user with a strong gain at BS1 has a weak gain at BS2. Profile 3 is an intermediate scenario where strong users at BS1 are weak at BS2 and moderate users at BS1 are strongest at BS2. This approach gives a limiting link gain profile as $K\to \infty $ and allows us to investigate convergence. However, it is tightly constrained by the choice of $\beta (x)$ and the Profiles in Figure \ref{LinkGainProfiles}. Hence, it is awkward to construct reasonable scenarios for more than two antenna clusters due to the proliferation of potential profiles and this approach is only used to illustrate convergence for $N=1$ and $N=2$.
\subsection{Imperfect CSI Model}
\label{ImCSIModel}
The estimated channel matrix, $\hat{\bG}$, in an imperfect CSI scenario is given by \cite{SURAWEERA}
\begin{equation} \label{impCSIG}
	\hat{\bG} = \xi \bG+\sqrt{1-\xi ^{2}}\bE,
\end{equation}
where $\bE$ is independent and statistically identical to $\bG$ and $0\leq \xi \leq 1$ controls the accuracy of the CSI.
\subsection{Correlation Model}
\label{CorrModel}
As we increase the density of the antennas in a cluster, the correlation among antenna elements will usually increase. Here, the correlation between antenna elements in an antenna cluster is modeled using the simple exponential model \cite{SMITH}, 
\begin{equation}
	\rho _{ij}= a^{-d_{ij}} \label{corr_rho} ,
\end{equation}
where $d_{ij}$ is the distance between the $i$th and $j$th pair of x-pol antennas and $0<a<1$. The $\frac{M}{N}\times \frac{M}{N}$ transmit correlation matrix, $\bR_{\textrm{t}}$, for each antenna cluster is modeled by the Kronecker structure,
\begin{equation} \label{CorrMatrix}
	\bR_{\textrm{t}} = \begin{bmatrix}	
		\bX_{\textrm{pol}} & \rho _{1,2}\bX_{\textrm{pol}} &	\ldots 	& \rho _{1,\frac{M}{2N}}\bX_{\textrm{pol}} \\
       	\rho _{2,1}\bX_{\textrm{pol}} & \bX_{\textrm{pol}} &	\ldots 	& \rho _{2,\frac{M}{2N}}\bX_{\textrm{pol}} \\
		\vdots & \vdots & \ddots & \vdots \\
       	\rho _{\frac{M}{2N},1}\bX_{\textrm{pol}} & \rho _{\frac{M}{2N},2}\bX_{\textrm{pol}}	& \ldots & \bX_{\textrm{pol}}
     	\end{bmatrix},
\end{equation}
where $\bX_{\textrm{pol}}$ is the $2\times 2$ x-pol antenna matrix given by
\begin{equation}
	\bX_{\textrm{pol}} = \begin{bmatrix}
					1 & r_{\textrm{pol}} \\
					r_{\textrm{pol}} & 1 
				\end{bmatrix},
\end{equation}
with $r_{\textrm{pol}}$ denoting the correlation between the two antenna elements in the x-pol pair. A fixed size  array is considered, where the x-pol antennas are positioned in a square shaped configuration.
%
%
\section{SINR Analysis}
\label{SINRAnalysis}
\subsection{Preliminaries}
We begin by outlining several mathematical results which will be used in the subsequent sections. 
\par
$1)$ Denote the $i$th column of $\bG $ as $\bg_{i} = \bP_{i}^{\frac{1}{2}}\bu_{i}$, where $\bu_{i} \in \mathbb{C}^{M\times 1}$ contains independent $\mathcal{CN}(0,1)$ elements. Note that the $M\times M$ matrix $\bP_{i}^{\frac{1}{2}}$ contains both the link gain coefficients and spatial correlation effects. Since, in \eqref{impCSIG}, $\bE$ is statistically identitical to $\bG$, we can write
\begin{align}
	\mathbb{E}\left[ \be_{i}^{\ast }\be_{i}^{\textrm{T}}\right] 
	&= \mathbb{E}\left[ \bg_{i}^{\ast }\bg_{i}^{\textrm{T}}\right] \\
	&= \mathbb{E}\left\{ 
		   \begin{bmatrix}
  			   	\beta _{1i}^{\frac{1}{2}}\bR_{\textrm{t}}^{\frac{1}{2}}\bH_{1i}^{\ast } \\
  			   	\vdots \\ 
  			   	\beta _{Ni}^{\frac{1}{2}}\bR_{\textrm{t}}^{\frac{1}{2}}\bH_{Ni}^{\ast }
	 	   \end{bmatrix}
	 	    \begin{bmatrix}
  			   	\beta _{1i}^{\frac{1}{2}}\bH_{1i}^{\textrm{T}}\bR_{\textrm{t}}^{\frac{1}{2}} &
  			   	\ldots &
  			   	\beta _{Ni}^{\frac{1}{2}}\bH_{Ni}^{\textrm{T}}\bR_{\textrm{t}}^{\frac{1}{2}}
	 	   \end{bmatrix}
	 	   \right\} \label{prelim1_start} 
\end{align}
\begin{align}
	&= \begin{bmatrix}
  			   	\beta _{1i}\bR_{\textrm{t}} & & \\
  			   	& \ddots & \\ 
  			   	& & \beta _{Ni}\bR_{\textrm{t}}
	   \end{bmatrix} 
	= \bP_{i}, \label{prelim1_end}
\end{align}
where $\be_{i}$ is the $i$th column of $\bE$. 
\par
$2)$ Using the eigen-decomposition of the transmit correlation matrix, $\bR_{\textrm{t}}$, we have from \eqref{prelim1_end}
\begin{align}
	\bP_{i} &= 
	 	 \begin{bmatrix}
  			   	\beta _{1i}\boldsymbol\psi ^{\textrm{T}}\boldsymbol\Lambda \boldsymbol\psi ^{\ast}  & &	\\
  				& \ddots &	\\
  				& & \beta _{Ni}\boldsymbol\psi ^{\textrm{T}}\boldsymbol\Lambda \boldsymbol\psi ^{\ast} 
	 	   \end{bmatrix}  \label{prelim2_start} \\
	 	&= \begin{bmatrix}
  			   	\boldsymbol\psi ^{\textrm{T}} & &	\\
  				& \ddots &	\\
  				& & \boldsymbol\psi ^{\textrm{T}} 
	 	   \end{bmatrix}
	 	   \begin{bmatrix}
  			   	\beta _{1i}\boldsymbol\Lambda  & &	\\
  				& \ddots &	\\
  				& & \beta _{Ni}\boldsymbol\Lambda 
	 	   \end{bmatrix}
	 	   \begin{bmatrix}
  			   	\boldsymbol\psi ^{\ast}  & &	\\
  				& \ddots &	\\
  				& & \boldsymbol\psi ^{\ast} 
	 	   \end{bmatrix} \\
	 	&= \boldsymbol\phi ^{\textrm{T}} \bQ_{i} \boldsymbol\phi ^{\ast }, \label{prelim2_end}
\end{align}
where $\boldsymbol\Lambda $ is the diagonal matrix of eigenvalues of $\bR_{\textrm{t}}$ and $\boldsymbol\psi $ is unitary. Similarly, $\bQ_{i}$ is a diagonal matrix containing the eigenvalues of $\bP_{i}$ and $\boldsymbol\phi$ is unitary. Note that $\boldsymbol\phi$ is fixed for all $\bP_{i}$, as it only depends on $\bR_{\textrm{t}}$, which we assume to be the same at each antenna cluster. 
\par
$3)$ For $\bv \in \mathbb{C}^{M\times 1}$ with independent $\mathcal{CN}(0,1)$ elements, and, for an arbitrary $\bP _{i}$ we show in Appendix \ref{prelim3_deriv} that
\begin{equation}
	\lim_{K\to \infty}\mathbb{E}\left[ \frac{1}{M}\bv^{\textrm{T}}\bP_{i} \bv^{\ast }\right] = \overline{\beta _{i}}, \label{prelim3}
\end{equation}
where $\overline{\beta _{i}}$ is the average of $\beta _{1i},\beta _{2i},\ldots ,\beta _{Ni}$. 
%
%
%
%
%
\subsection{MF Precoding}
\label{MFPrecoding}
%
Having outlined the prerequisite mathematical results, we now derive the limiting SINR expressions for MF precoding with large distributed antenna arrays.
\par
The transmitted signal for a MF precoder, with CSI inaccuracy, is given by
\begin{equation}
	\bs = \frac{1}{\sqrt{\gamma }}\hat{\bG}^{\ast }\bq,
\end{equation}
with $\mathbb{E}\left[ \| \bq\| ^{2}\right] =1$, where $\bq$ is the $K\times 1$ data symbol vector, and the average power is normalized by
\begin{equation}
	\gamma = \frac{\textrm{tr}(\hat{\bG}^{\textrm{T}}\hat{\bG}^{\ast })}{K}.
\end{equation}
The combined received signal for all users is thus given by
\begin{equation} \label{mf_x}
	\bx = \sqrt{\frac{\rho _{\textrm{f}}}{\gamma }}\bG^{\textrm{T}}\hat{\bG}^{\ast }\bq + \bw,
\end{equation}
with the $i$th user receiving the $i$th component of $\bx $, given by
\begin{equation} \label{mf_xi}
	x_{i} = \sqrt{\frac{\rho _{\textrm{f}}}{\gamma }}\bg_{i}^{\textrm{T}}\hat{\bG}^{\ast }\bq + w_{i}.
\end{equation}
The expected value of the power of the $i$th users received signal in \eqref{mf_xi} can be shown to be (see Appendix \ref{mf_sig_power})
\begin{align}
	\mathbb{E}\left[ P_{\textrm{sig},i}\right] 
	&= \mathbb{E}\left[ \left| \sqrt{\frac{\rho _{\textrm{f}}}{\gamma }}\bg_{i}^{\textrm{T}}\hat{\bg}_{i}^{\ast }q_{i} \right| ^{2}\right] \\
	&= \frac{\rho _{\textrm{f}}}{K\gamma }\left( \xi ^{2}|\hat{\bg}_{i}^{\textrm{T}}\hat{\bg}_{i}^{\ast }|^{2} +(1-\xi ^{2}) \hat{\bg}_{i}^{\textrm{T}}\bP_{i}\hat{\bg}_{i}^{\ast }\right) \label{Psig},
\end{align}
where $\hat{\bg}_{i}$ is the $i$th column of $\hat{\bG}$. Likewise, the expected value of the interference and noise power of the $i$th user's received signal is (see Appendix \ref{mf_int_power})
\begin{align}
	\mathbb{E}\left[ P_{\textrm{i+n},i}\right] 
	&= \mathbb{E}\left[ \left| \sqrt{\frac{\rho _{\textrm{f}}}{\gamma }}\sum_{k=1,k\neq i}^{K}{\bg_{i}^{\textrm{T}}\hat{\bg}_{k}^{\ast }q_{k}+w_{i}}\right| ^{2}\right] \\
	&= \frac{\rho _{\textrm{f}}}{K \gamma }\sum_{k=1,k\neq i}^{K}{\left( \xi ^{2} \left| \hat{\bg}_{i}^{\textrm{T}}\hat{\bg}_{k}^{\ast }\right| ^{2} +(1-\xi ^{2}) \hat{\bg}_{k}^{\textrm{T}}\bP_{i}\hat{\bg}_{k}^{\ast } \right) } + \sigma ^{2} \label{Pin}.
\end{align}
Combining \eqref{Psig} and \eqref{Pin}, the expected MF SINR for the $i$th user is given by \cite{YU}
\begin{equation} \label{mf_sinr}
	\mathbb{E}\left[ \textrm{SINR}_{i}\right] \approx \frac{
	\frac{\rho _{\textrm{f}}}{K\gamma }\left( \xi ^{2}\left| \hat{\bg}_{i}^{\textrm{T}}\hat{\bg}_{i}^{\ast }\right| ^{2} +(1-\xi ^{2}) \hat{\bg}_{i}^{\textrm{T}}\bP_{i}\hat{\bg}_{i}^{\ast } \right)
	}
	{
	\frac{\rho _{\textrm{f}}}{K\gamma }\sum\limits_{k=1,k\neq i}^{K}{\left( \xi ^{2}\left| \hat{\bg}_{i}^{\textrm{T}}\hat{\bg}_{k}^{\ast }\right| ^{2} +(1-\xi ^{2}) \hat{\bg}_{k}^{\textrm{T}}\bP_{i}\hat{\bg}_{k}^{\ast } \right) } + \sigma ^{2}
	}.
\end{equation}
We wish to study the asymptotic behaviour of \eqref{mf_sinr} given by
\begin{equation} 
	\lim_{K\to \infty }\mathbb{E}\left[ \textrm{SINR}_{i}\right]  
	= \frac{
	\frac{\rho _{\textrm{f}}\alpha }{\lim\limits_{K\to \infty } \left( \frac{\gamma }{M}\right) }\left( \xi ^{2}\lim\limits_{K\to \infty }\left|  \frac{\hat{\bg}_{i}^{\textrm{T}}\hat{\bg}_{i}^{\ast }}{M}\right| ^{2} + \lim\limits_{K\to \infty }\left( \frac{1-\xi ^{2}}{M} \frac{\hat{\bg}_{i}^{\textrm{T}}\bP_{i}\hat{\bg}_{i}^{\ast }}{M} \right) \right)
	}{
	\sigma ^{2} + \frac{\rho _{\textrm{f}}\alpha }{\lim\limits_{K\to \infty } \left( \frac{\gamma }{M}\right) }\lim\limits_{K\to \infty }\sum\limits_{k=1,k\neq i}^{K}{\left( \xi ^{2} \left| \frac{\hat{\bg}_{i}^{\textrm{T}}\hat{\bg}_{k}^{\ast }}{M}\right| ^{2} + \frac{(1-\xi ^{2})}{M} \frac{\hat{\bg}_{k}^{\textrm{T}}\bP_{i}\hat{\bg}_{k}^{\ast }}{M} \right)}
	} .\label{mf_sinr_2} 
\end{equation}
In the numerator, since $\hat{\bg}_{i}$ has the same statistics as $\bg_{i}$, we can write $\hat{\bg}_{i} = \bP_{i}^{\frac{1}{2}}\bv_{i} $ where the elements of $\bv_{i}$ are i.i.d. $\mathcal{CN}(0,1)$. Hence, $\hat{\bg}_{i}^{\textrm{T}}\hat{\bg}_{i}^{\ast } =  \bv_{i}^{\textrm{T}}\bP_{i}^{\frac{1}{2}}\bP_{i}^{\frac{1}{2}}\bv_{i}^{\ast } = \bv_{i}^{\textrm{T}}\bP_{i}\bv_{i}^{\ast }$. Then, using \eqref{prelim3}, we have
\begin{equation}
	\lim_{K\to \infty} \frac{1}{M}\hat{\bg}_{i}^{\textrm{T}}\hat{\bg}_{i}^{\ast } = \lim_{K\to \infty }\frac{1}{N}\sum_{n=1}^{N}{\beta _{ni}} = \overline{\beta _{i}}.
\end{equation}
Similarly, $\hat{\bg}_{i}^{\textrm{T}}\bP_{i}\hat{\bg}_{i}^{\ast }= \bv_{i}^{\textrm{T}}\bP_{i}^{2}\bv_{i}^{\ast }$ and a simple extension of \eqref{prelim3} gives
\begin{align}
	\lim_{K\to \infty }\frac{1}{M}\hat{\bg}_{i}^{\textrm{T}}\bP_{i}\hat{\bg}_{i}^{\ast }
	&= \lim_{K\to \infty }\left( \frac{1}{N}\sum_{n=1}^{N}{\beta _{ni}^{2}}\right) \lim_{K\to \infty }\left( \sum_{i=1}^{M/N}{\frac{\Lambda _{ii}^{2}}{M/N}}\right) \\
	&= \overline{\beta _{i}^{2}}~\overline{\Lambda ^{2}}, \label{28}
\end{align}
where $\overline{\beta _{i}^{2}}$ is the average of $\beta _{1i}^{2},\beta _{2i}^{2},\ldots ,\beta _{Ni}^{2}$ and $\overline{\Lambda ^{2}}$ is the limiting average of \\$\Lambda _{11}^{2},\Lambda _{22}^{2},\ldots ,\Lambda _{\frac{M}{N}\frac{M}{N}}^{2}$. Also,
\begin{align}
	\lim_{K\to \infty} \frac{\gamma }{M}
	&= \lim_{K\to \infty}\frac{1}{M} \frac{\textrm{tr}(\hat{\bG}^{\textrm{T}}\hat{\bG}^{\ast })}{K} \\
	&= \lim_{K\to \infty }\frac{1}{K}\sum_{i=1}^{K}{\frac{1}{M}\hat{\bg}_{i}^{\textrm{T}}\hat{\bg}_{i}^{\ast } } \\
	&= \lim_{K\to \infty }\frac{1}{NK}\sum_{n=1}^{N}{\sum_{k=1}^{K}{\beta _{nk}}} \\
	&= \overline{\beta },
\end{align}
where $\overline{\beta }$ is the limiting average of the $\beta _{n,k}$ values over $n$ and $k$. Since the limit in \eqref{28} is finite and $\frac{1-\xi ^{2}}{M}\rightarrow 0$ as $K\to \infty $, the final limit term in the numerator of \eqref{mf_sinr_2} approaches zero. Therefore, the asymptotic limit of the numerator of \eqref{mf_sinr_2} is given as
\begin{equation}
	\frac{\rho _{\textrm{f}}\alpha }{\lim\limits_{K\to \infty } \left( \frac{\gamma }{M}\right) }\left( \xi ^{2}\lim\limits_{K\to \infty }\left|  \frac{\hat{\bg}_{i}^{\textrm{T}}\hat{\bg}_{i}^{\ast }}{M}\right| ^{2} + \lim\limits_{K\to \infty }\left( \frac{1-\xi ^{2}}{M} \frac{\hat{\bg}_{i}^{\textrm{T}}\bP_{i}\hat{\bg}_{i}^{\ast }}{M}\right) \right) 
	= \rho _{\textrm{f}}\alpha \frac{1}{\overline{\beta }}\xi ^{2}\overline{\beta _{i}}^{2}. \label{asym_Psig}
\end{equation}
Likewise, we examine the denominator of \eqref{mf_sinr_2} as $K\rightarrow \infty $. It is shown in Appendix \ref{int_lim_1} that
\begin{equation}
	\lim_{K\rightarrow \infty }\frac{1}{M^2}\sum_{k=1,k\neq i}^{K}{\left| \hat{\bg}_{i}^{\textrm{T}}\hat{\bg}_{k}^{\ast }\right| ^{2} }
	= \frac{1}{\alpha }\overline{\Lambda ^{2}}~\overline{\beta _{ik}},
\end{equation}
where $\overline{\beta _{ik}}$ is the limiting average cross product of the $i$th user's link gains with all the other users' link gains. Also in Appendix \ref{int_lim_2} it is shown that
\begin{equation}
	\lim_{K\to \infty}\sum_{k=1,k\neq i}^{K}{ \frac{\hat{\bg }_{k}^{\textrm{T}}\bP_{i}\hat{\bg }_{k}^{\ast }}{M^{2}}} = \frac{1}{\alpha }\overline{\Lambda ^{2}}~\overline{\beta _{ik}}.
\end{equation}
Therefore, the asymptotic limit of the denominator of \eqref{mf_sinr_2} is given as
\begin{align}
	&\sigma ^{2} + \frac{\rho _{\textrm{f}}\alpha }{\lim\limits_{K\to \infty } \left( \frac{\gamma }{M}\right) }\lim\limits_{K\to \infty }\sum\limits_{k=1,k\neq i}^{K}{\left( \xi ^{2} \left| \frac{\hat{\bg}_{i}^{\textrm{T}}\hat{\bg}_{k}^{\ast }}{M}\right| ^{2} + \frac{(1-\xi ^{2})}{M} \frac{\hat{\bg}_{k}^{\textrm{T}}\bP_{i}\hat{\bg}_{k}^{\ast }}{M} \right)} \notag \\
	&= \sigma ^{2} + \frac{\rho _{\textrm{f}}\alpha }{\overline{\beta }}\left( \xi ^{2}\frac{1}{\alpha }\overline{\Lambda ^{2}}~\overline{\beta _{ik}}+(1-\xi ^{2})\frac{1}{\alpha }\overline{\Lambda ^{2}}~\overline{\beta _{ik}}\right) \\
	&= \sigma ^{2}+\frac{\rho _{\textrm{f}}\overline{\Lambda ^{2}}~\overline{\beta _{ik}}}{\overline{\beta }}. \label{asym_Pin}
\end{align}
Substituting \eqref{asym_Psig} and \eqref{asym_Pin} into \eqref{mf_sinr_2} gives the limit of the expected per-user MF SINR expression \eqref{mf_sinr} as
\begin{equation} 
	\lim_{K\rightarrow \infty }\textrm{SINR}_{i} = \frac{\rho _{\textrm{f}}\alpha \xi ^{2}\overline{\beta _{i}}^{2}}{\overline{\beta }+\rho _{\textrm{f}}\overline{\beta _{ik}}~\overline{\Lambda ^{2}}}, \label{mf_sinr_lim}
\end{equation}
where the noise power is normalized to 1. 
\par
From \eqref{mf_sinr_lim}, we can examine the effects of each component on the SINR limit. The transmit SNR, $\rho _{\textrm{f}}$, boosts the signal power but also the interference power leading to a ceiling on the SINR limit, as $\textrm{SINR}_{i}\rightarrow \frac{\alpha \xi ^{2}\overline{\beta _{i}}^{2}}{\overline{\beta _{ik}}~\overline{\Lambda ^{2}}}$ as $K\to \infty ,\rho _{\textrm{f}}\to \infty $. The ratio, $\alpha $, increases the SINR due to increased diversity. The CSI factor, $\xi $, decreases the signal power but the extra interference created by imperfect CSI disappears in the limit due to averaging. $\overline{\Lambda ^{2}}$ reduces the SINR and implies that correlation reduces SINR. To see this, consider the extreme cases of an i.i.d. channel ($\bR_{\textrm{t}}=\Lambda =\Lambda ^{2}=\bI_{M/N}$) and a fully correlated channel ($\bR_{\textrm{t}}=\mathbf{1} _{M/N}$, $\Lambda =\textrm{diag}(M/N,0,0,\ldots ,0)$, $\Lambda ^{2}=\textrm{diag}((M/N)^{2},0,0,\ldots ,0)$), where $\mathbf{1} _{M/N}$ is an $M/N\times M/N$ matrix of ones. These scenarios give $\overline{\Lambda ^{2}}=1$ and $\overline{\Lambda ^{2}}=M/N$, respectively. Clearly, the $\overline{\Lambda ^{2}}$ term increases with correlation and reduces the SINR limit. $\overline{\beta }$ reduces performance as it is a measure of the total power of the received signals which includes the aggregate interference. $\overline{\beta _{ik}}$ reduces performance as it is an inverse measure of orthogonality. If the desired user $i$ has strong links on the antennas in a set of clusters $\mathcal{A}\subset \{ 1,2,\ldots ,N\} $ and all the interferers have weak link gains in $\mathcal{A}$ then the ``cross product'' term $\overline{\beta _{ik}}$ is weak. Here, the channels are close to orthogonal (on average) and performance is enhanced.
\par
Note that in \eqref{mf_sinr_lim}, we have presented a per-user limiting value which can be evaluated for a particular link gain model. As an example, we evaluate the per-user SINR using the limiting link gain model described in Section \ref{LinkGainModel}. Without loss of generality, we consider the co-located ($N=1$) BS case where the link gain profile is decaying (exponentially). Thus, evaluating the terms in \eqref{mf_sinr_lim}, we have
\begin{equation}
	\overline{\beta _{i}} = \frac{1}{N}\sum_{n=1}^{N}{\beta _{ni}} = \beta _{1i},
\end{equation}
and
\begin{align}
	\overline{\beta } &= \lim_{K\to \infty}\frac{1}{NK}\sum_{n=1}^{N}{\sum_{k=1}^{K}{\beta _{nk}}} \\
	&= \lim_{K\to \infty }\frac{1}{K}\sum_{k=1}^{K}{\beta _{1k}} \\
	&= \int_{0}^{1}{\beta (x)dx} \\
	&= \int_{0}^{1}{\beta _{\textrm{max}}(\beta _{\textrm{min}}/\beta _{\textrm{max}})^{x}dx} \\
	&= \frac{\beta _{\textrm{max}}-\beta _{\textrm{min}}}{\textrm{log}_{\textrm{e}}(\beta _{\textrm{max}})-\textrm{log}_{\textrm{e}}(\beta _{\textrm{min}})}, \label{tmp4}
\end{align}
where \eqref{tmp4} follows from standard methods. Also,
\begin{align}
	\overline{\beta _{ik}} &= \lim_{K\to \infty}\frac{1}{NK}\sum_{n=1}^{N}{\sum_{k=1,k\neq i}^{K}{\beta _{ni}\beta _{nk}}} \\
	&= \lim_{K\to \infty}\frac{1}{K}\sum_{k=1,k\neq i}^{K}{\beta _{1i}\beta _{1k}} \\
	&= \beta _{1i}\overline{\beta } 
\end{align}
\begin{align}
	&= \beta _{1i}\frac{\beta _{\textrm{max}}-\beta _{\textrm{min}}}{\textrm{log}_{\textrm{e}}(\beta _{\textrm{max}})-\textrm{log}_{\textrm{e}}(\beta _{\textrm{min}})}.
\end{align}
Hence, for any limiting link gain model, the exact limit in \eqref{mf_sinr_lim} can be evaluated.
\par
Finally, we can consider several special cases of \eqref{mf_sinr_lim}:
\subsubsection{Perfect CSI}
\begin{equation}
	\lim_{K\rightarrow \infty }\textrm{SINR}_{i} = \frac{\rho _{\textrm{f}}\alpha \overline{\beta _{i}}^{2}}{\overline{\beta } + \rho _{\textrm{f}}\overline{\beta _{ik}}~ \overline{\Lambda ^{2}}}.
\end{equation}
\subsubsection{No Spatial Correlation}
\begin{equation}
	\lim_{K\rightarrow \infty }\textrm{SINR}_{i} = \frac{\rho _{\textrm{f}}\alpha \xi ^{2}\overline{\beta _{i}}^{2}}{\overline{\beta } + \rho _{\textrm{f}}\overline{\beta _{ik}}} \label{mf_sinr_lim_no_corr}.
\end{equation}
\subsubsection{Equal Power Distribution with Spatial Correlation}
\begin{equation}
	\lim_{K\rightarrow \infty }\textrm{SINR}_{i} = \frac{\rho _{\textrm{f}}\alpha \xi ^{2}\beta }{1 + \rho _{\textrm{f}}\beta \overline{\Lambda ^{2}}}.
\end{equation}
Here, the link gain for all users from all clusters is a constant, $\beta $.
\subsubsection{No Spatial Correlation, Equal Power Distribution}
\begin{equation}
	\lim_{K\rightarrow \infty }\textrm{SINR}_{i} = \frac{\rho _{\textrm{f}}\alpha \xi ^{2}\beta }{1 + \rho _{\textrm{f}}\beta } \label{rusek1}.
\end{equation}
\subsubsection{No Spatial Correlation, Equal Power Distribution, Perfect CSI}
\begin{equation} 
	\lim_{K\rightarrow \infty }\textrm{SINR}_{i} = \frac{\rho _{\textrm{f}}\alpha \beta }{1 + \rho _{\textrm{f}}\beta } \label{rusek2}.
\end{equation}
Note that \eqref{rusek1} and \eqref{rusek2} agree with the results given in \cite{RUSEK} when $\beta =1$. 
\section{Numerical Results}
\label{NumericalResults}
%
\subsection{Simulation Parameters}
\label{simParameters}
In Section \ref{convergence}, we illustrate the convergence of the mean per-user SINR to its limiting expression, for $N=1$ and $2$ BSs, using the limiting link gain model described in Section \ref{LinkGainModel}. In Sections \ref{uncorr} and \ref{corr}, the uncorrelated and correlated performance of MF precoding is shown for $N\in \left\{ 1,2,5\right\} $ BSs, using the statistical link gain model described below.
\par
For the limiting link gain model, values of $\rho _{\textrm{f}}\beta _{\textrm{max}}=25 \textrm{ dB}$ and $\rho _{\textrm{f}}\beta _{\textrm{min}}=-5 \textrm{ dB}$, where $\rho _{\textrm{f}} =10 \textrm{ dB}$, are arbitrarily chosen. For all simulations after Section \ref{convergence}, the statistical link gain model is used. We calculate the path-loss between each user and the BSs using $ALd^{-\gamma }$, where $L$ is log-normal shadowing and $d$ is the link distance. The shadow fading standard deviation is $\sigma =8$ dB, the path-loss exponent is $\gamma =4$ and the link distance is $50<d<1000\textrm{ m}$, unless otherwise stated. $A$ is used as an offset, such that the maximum link gain generated from the statistical link gain model aligns with that of the maximum limiting link gain model value, i.e., $A=\frac{\beta _{\textrm{max}}}{\textrm{max}(Ld^{-\gamma })}$. 
\par
For simulations involving a single, co-located, antenna cluster, we position the BS in the center of the coverage region, whereas, in simulations considering $N\geq 2$ BSs, the antenna clusters are positioned equidistant on the periphery of the coverage region. The exponential model correlation parameter in \eqref{corr_rho} is arbitrarily chosen to be $a=4$ (another value could be chosen). The correlation matrix, given in \eqref{CorrMatrix}, is calculated using a carrier frequency of $2.6 \textrm{ GHz}$ and a $1 \times 1 \textrm{ m}$ square antenna array, with the antenna correlation between two elements in the same x-pol configuration, $r_{\textrm{pol}}=0.1$. 
All results are simulated with $\alpha =\frac{M}{K}=10$ and $\rho _{\textrm{f}}=10$ dB.
\subsection{Convergence of $N=1$ and $2$ BSs}
\label{convergence}
\begin{figure}[ht]
\centering\includegraphics[width=110mm]{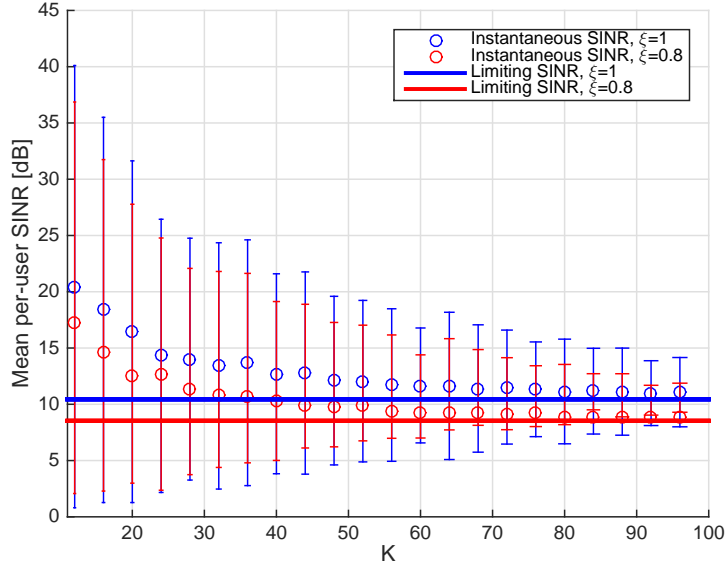}
\raisecaption\caption{Instantaneous and limiting MF SINR as a function of the number of users, $K$, for a co-located antenna cluster, $N=1$, with no transmit spatial correlation, $\overline{\Lambda ^{2}}=1$.}
\label{convergence_1bs}
\end{figure}
\begin{figure}[ht]
\centering\includegraphics[width=110mm]{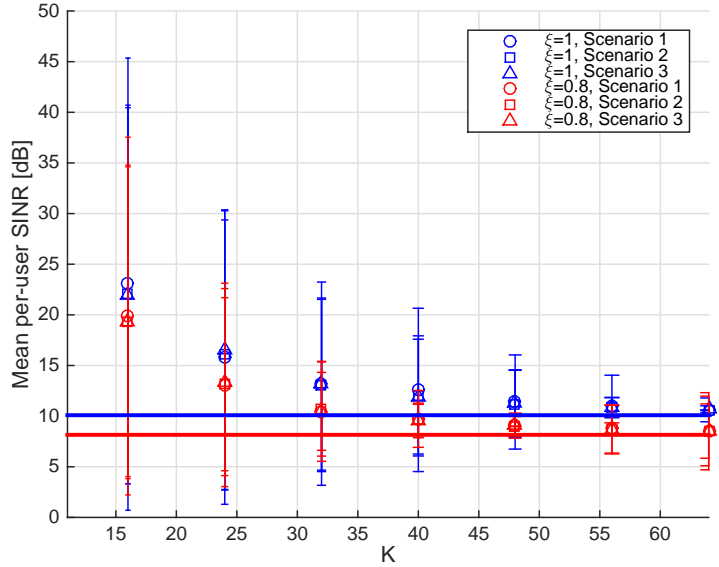}
\raisecaption\caption{Instantaneous and limiting MF SINR as a function of the number of users, $K$, for two antenna clusters, $N=2$, each with no transmit spatial correlation, $\overline{\Lambda ^{2}}=1$.}
\label{convergence_2bs}
\end{figure}
We begin by illustrating the convergence of mean per-user SINR to the corresponding limiting value, for $N=1$ and $N=2$ BSs, where we use the limiting link gain model, outlined in Section \ref{LinkGainModel}.
\par
Figures \ref{convergence_1bs} and \ref{convergence_2bs} show the convergence of mean per-user SINR (i.e., averaged over all users), given in \eqref{mf_sinr}, to its mean per-user limiting value, \eqref{mf_sinr_lim}, for one and two BSs respectively. Note that the points are the instantaneous SINRs averaged across the $K$ users and over the fast fading. The bars above and below the points represent the plus/minus one standard deviation limits of the instantaneous SINRs averaged over the users. As the width of the error bars decreases with $K$, we see that the instantaneous SINRs converge to the limit in addition to the expected SINR (over fast fading). In comparing the two figures, we observe that the additional BS has almost no effect on both the rate of convergence and mean per-user SINR for large systems. This is due to the fact that $\overline{\beta _{ik}}$ in the limiting SINR expression, \eqref{mf_sinr_lim}, tends to be small compared to $\overline{\beta }$. In both cases the mean per-user SINR has effectively reached its limiting value, of approximately $10$ dB for perfect CSI, for a system of size $K=100$ single antenna users, i.e., 1000 total BS antennas. The effect of BS numbers on both mean per-user SINR and rate of convergence is explored more thoroughly for a larger number of BSs in later results. It can also be seen that the reduction in CSI results in a decrease in the mean limiting per-user SINR of about 2 dB in both cases. This is due to the linear relationship between CSI imperfections and limiting per-user SINR, shown in \eqref{mf_sinr_lim}. 
\par
Given the results in Figure \ref{convergence_2bs}, we conclude that rate of convergence and performance of both the limiting and simulated mean per-user SINR is largely independent of the link gain profile used, outlined in Section \ref{LinkGainModel}, used to generate the users' link gains. 
For larger numbers of antennas, the profiles have little effect and this is consistent with the fact that with MF, the aggregate interference is the dominant factor.
\subsection{Uncorrelated MF SINR Performance}
\label{uncorr}
\begin{figure}[ht]
\centering\includegraphics[width=110mm]{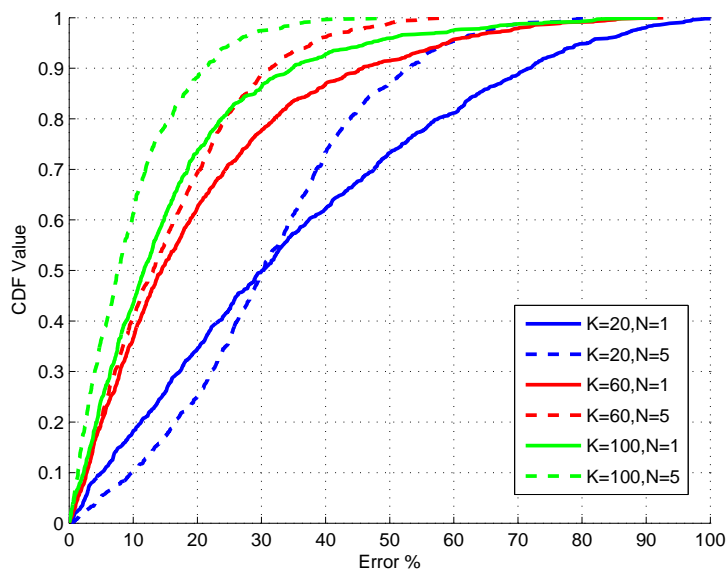}
\raisecaption\caption{Instantaneous MF SINR error $\% $ CDF as a function of the number of users, $K$, and antenna clusters, $N$. Here, we have perfect CSI, $\xi =1$, and no transmit spatial correlation.}
\label{error_cdf_uncorr}
\end{figure}
\begin{figure}[ht]
\centering\includegraphics[width=110mm]{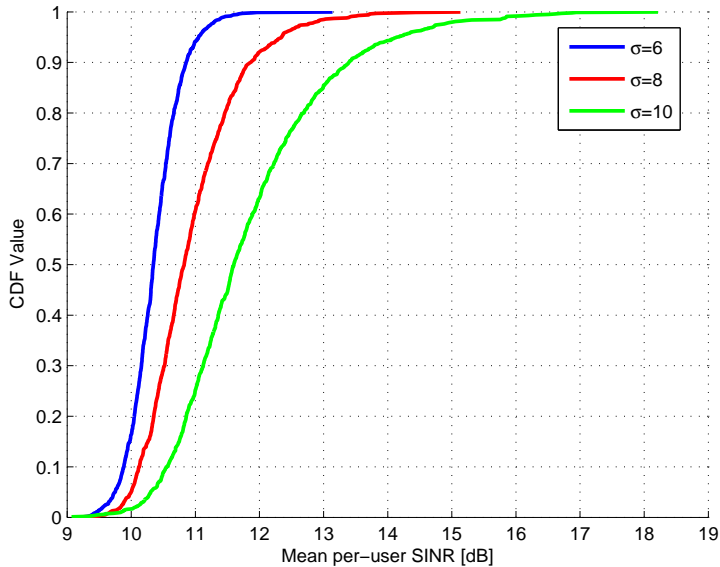}
\raisecaption\caption{Instantaneous MF SINR CDF as a function of the shadowing variance, $\sigma $, for $K=60$ and $N=5$. Here, we have perfect CSI, $\xi =1$, and no transmit spatial correlation.}
\label{shad_var_uncorr}
\end{figure}
\begin{figure}[ht]
\centering\includegraphics[width=110mm]{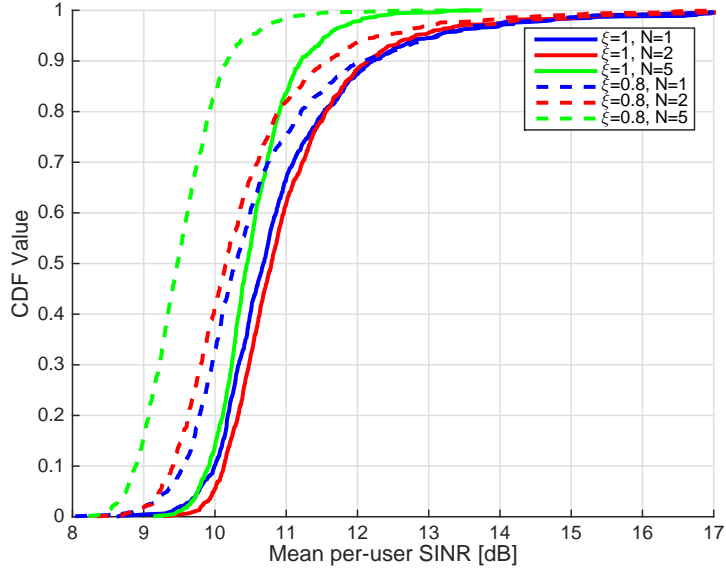}
\raisecaption\caption{Instantaneous MF SINR CDF as a function of CSI imperfections, $\xi $, and antenna clusters numbers, $N$. Here, we have $K =100$, and no transmit spatial correlation.}
\label{sinr_cdf_uncorr}
\end{figure}
\begin{figure}[ht]
\centering\includegraphics[width=110mm]{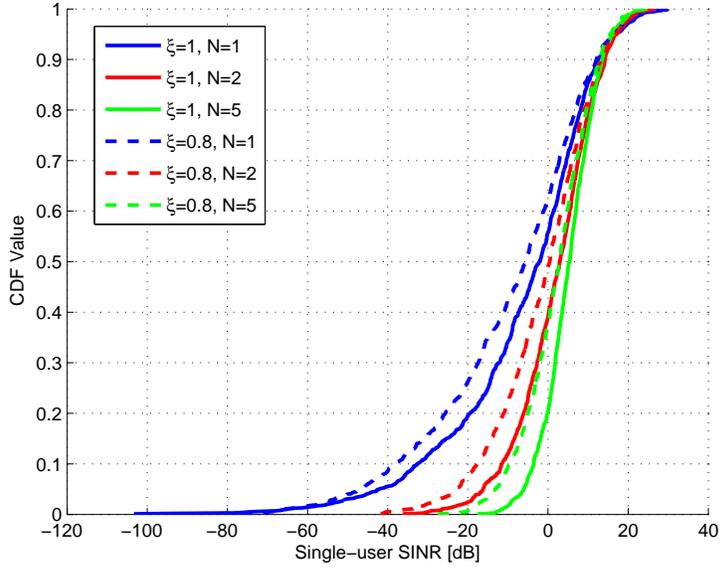}
\raisecaption\caption{Single-user MF SINR CDF as a function of CSI imperfections, $\xi $, and antenna clusters numbers, $N$. Here, we have $K =100$, and no transmit spatial correlation.}
\label{su_uncorr_sinr}
\end{figure}

We now illustrate the uncorrelated performance of MF precoders, where the statistical link gain model is used to generate users' link gains is used. In Figures \ref{error_cdf_uncorr}-\ref{sinr_cdf_corr} we compute the instantaneous SINRs averaged over the $K$ users for many independent drops. Note that each drop has a different set of link gains and therefore a different limiting SINR. In Figure \ref{error_cdf_uncorr} we show how quickly the mean per-user SINR converges towards its mean per-user limiting value for different size systems in an uncorrelated scenario. The virtual limit is computed by simulating a system size of $M=1400$ BS antennas. We define: Error $\% = \frac{\left| \overline{\lim\limits_{K\to \infty }\textrm{SINR}}-\overline{\textrm{SINR}}\right| }{\overline{\textrm{SINR}}}\times 100$, where $\overline{\lim\limits_{K\to \infty }\textrm{SINR}}$ is the mean per-user SINR limit and $\overline{\textrm{SINR}}$ is the mean per-user SINR. We plot the Error $\% $ cumulative distribution function (CDF) for small, medium and large sized systems, corresponding to $K=20,60$ and $100$ single antenna users respectively. In each case of an increasing step in system size, by $K=40$, it can be seen that the change in Error $\% $ is reduced, e.g., for the median value for $N=5$ BSs, the error decreases by $\approx 30-12=18\% $ as we increase the system from small to medium size, whereas, the error decreases by $\approx 12-8=4\% $ from increasing the system size from $K=60$ to $100$ users. This decaying rate of the rate of convergence effect is also seen in Figures \ref{convergence_1bs} and \ref{convergence_2bs}.
\par
In Figure \ref{shad_var_uncorr}, we illustrate the impact of changing shadowing variance on the mean per-user SINR. A greater shadow variance is shown to produce larger SINR values and a greater SINR range, resulting from the increased variability in path-loss, as seen in the tails of each CDF. For example, the CDF with a shadowing variance of $\sigma =10$ has a range of SINRs from $18-9=9$ dB, in comparison to $13-9=4$ dB for $\sigma =6$.
\par
In Figure \ref{sinr_cdf_uncorr}, we show how both CSI imperfections and BS numbers impact the CDF of mean per-user SINR in an uncorrelated scenario. The single antenna cluster case outperforms the five BS case at a median value of $\approx 0.2$ dB and $\approx 1$ dB for perfect and imperfect transmitter channel knowledge respectively. The reason behind the co-located antenna array dominance is due to the underlying cell configuration used in simulation, described in Section \ref{simParameters}. For instance, for the $N=5$ case, if a user is close to a BS then it is receiving a strong signal from 200 antennas. Whereas, if a user is close to the BS in the $N=1$ scenario, it is receiving a strong signal from 1000 antennas, i.e., roughly speaking, the user is being served by an extra 800 degrees of freedom in the $N=1$ scenario. This is exemplified in the shape of the CDFs, where there is a large tail, at high SNR, for the co-located BS CDF. 
Furthermore, we notice that the larger BS cases do not perform better at low SNR, as we would expect. This is a result of the CDFs being a \emph{mean} of SINR across all users, rather than a single users CDF. Thus, we present a single-user uncorrelated SINR CDF, in Figure \ref{su_uncorr_sinr}, for comparison. 
\par
In Figure \ref{su_uncorr_sinr} we illustrate the MF SINR CDF for the single-user case. It can be seen that for low SNR, a larger number of BSs provides much better coverage, increasing SINR significantly. Despite the significant differences in the $N=1$ and $N=2$ case, there are smaller gains seen in increasing BS numbers from $N=2$ to $N=5$.
\subsection{Correlated MF SINR Performance}
\label{corr}
In this section, as with the uncorrelated MF SINR performance simulations in Section \ref{uncorr}, the statistical link gain model is used to generate the link gains for a scenario with spatial correlation.
\begin{figure}[ht]
\centering\includegraphics[width=110mm]{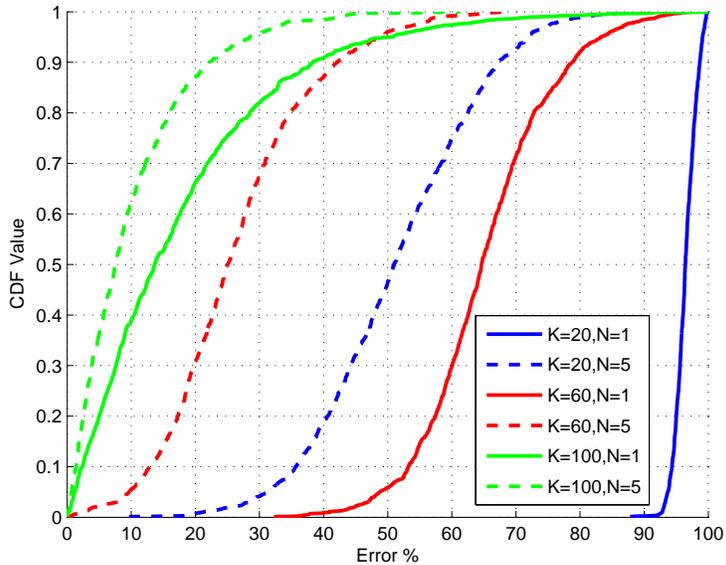}
\raisecaption\caption{Spatially correlated instantaneous MF SINR error $\% $ CDF as a function of the number of users, $K$, and antenna clusters, $N$. Here, we have perfect CSI, $\xi =1$.}
\label{error_corr}
\end{figure}
\par
In Figure \ref{error_corr}, we show the rate at which the mean per-user SINR approaches its limit in a correlated scenario, as a function of system size. As with the uncorrelated case, we consider three system sizes. It is clear that for a larger number of BSs, the mean SINR converges towards its limit much quicker than for a smaller number of BSs. The slower rate of convergence is due to the additional factor, $\overline{\Lambda ^{2}}$, in the piecewise convergence of \eqref{mf_sinr} to \eqref{mf_sinr_lim}, arising due to spatial correlation. In contrast with the uncorrelated error CDF, in Figure \ref{error_cdf_uncorr}, it is seen that correlation reduces the rate of convergence greatly for both cases for BS numbers. For example, the median Error $\% $ value for a medium sized system with a single antenna cluster is seen to increase by approximately $64-14=50\% $ when spatial correlation is introduced. On the other hand, the same scenario for $5$ antenna clusters, the impact of spatial correlation is shown to increase the Error $\% $ by approximately $25-13=12\%$. Thus, we see a large improvement in mitigating the impact of spatial correlation, on the rate of convergence, by distributing the antennas.
\begin{figure}[ht]
\centering\includegraphics[width=110mm]{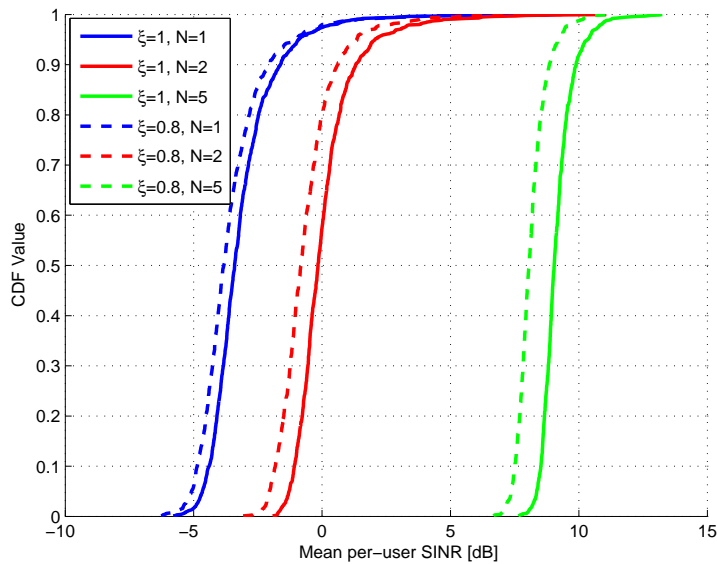}
\raisecaption\caption{Spatially correlated instantaneous MF SINR CDF as a function of CSI imperfections, $\xi $, and antenna clusters numbers, $N$. Here, we have $K =100$.}
\label{sinr_cdf_corr}
\end{figure}
\par
In Figure \ref{sinr_cdf_corr}, we show the impact of CSI imperfections and BS numbers on mean per-user SINR for a scenario with spatial correlation at the transmitter. As with the rate of convergence, when spatial correlation is present, we see a vast improvement in performance as we distribute the BS antennas into multiple clusters. This is more clearly seen by comparing the correlated and uncorrelated mean per-user SINR performances, given in Figures \ref{sinr_cdf_corr} and \ref{sinr_cdf_uncorr} respectively. In an uncorrelated scenario with CSI imperfections of $\xi =0.8$, the impact of increasing the BS numbers from $1$ to $5$ results in a loss of $\approx 10.7-9.4=1.3$ dB mean per-user SINR. Whereas, in a correlated scenario, we instead see a gain of $\approx 8-(-4)=12$ dB. Again, this is a result of $\overline{\Lambda ^{2}}$ in the denominator of \eqref{mf_sinr_lim}, which has a significantly negative effect when all antennas are co-located. To further quantify this effect, in Table \ref{corr_table} we tabulate $\overline{\Lambda ^{2}}$ for different numbers of BSs, $N$, and $r_{\textrm{pol}}$ values.
\begin{table}[h]
\centering
\begin{tabular}{|c|c|c|c|c|c|}
\hline
\multicolumn{1}{|l|}{} & \multicolumn{5}{c|}{$r_{\textrm{pol}}$}         \\ \hline
$N$             & \textbf{0.1} & \textbf{0.2} & \textbf{0.3} & \textbf{0.4} & \textbf{0.5} \\ \hline
\textbf{1}             & 28.71        & 29.57        & 30.99        & 32.98        & 35.54        \\ \hline
\textbf{2}             & 13.95        & 14.36        & 15.05        & 16.02        & 17.26        \\ \hline
\textbf{5}             & 1.42         & 1.46         & 1.53         & 1.63         & 1.75         \\ \hline
\textbf{10}            & 1.17         & 1.21         & 1.26         & 1.34         & 1.47         \\ \hline
\end{tabular}
\vspace{10pt}
\caption{$\overline{\Lambda ^{2}}$ values as a function of $N$ and $r_{\textrm{pol}}$}
\label{corr_table}
\end{table}
\par
Table \ref{corr_table} shows that, for all values of $r_{\textrm{pol}}$, we see huge improvements in the reduction of $\overline{\Lambda ^{2}}$ as we increase the number of BSs.
%
\section{Conclusion}
\label{sec:Conclusion}
In this paper, we have analyzed both the rate of convergence and the performance of a MF precoder in a massive MIMO system. We have presented a method to derive MF SINR for scenarios including: unequal link gains, imperfect CSI, transmitter spatial correlation and distributed BSs. From this, we have derived limiting expressions, as the number of antennas grow without bound, while considering several special cases.
\par
Results have shown that both the rate of convergence and precoder performance is largely dependent on the spatial correlation.  In the presence of spatial correlation, distributing of the antennas into multiple clusters renders significant gains over a co-located scenario. In uncorrelated scenarios, a co-located antenna cluster has a better mean per-user SINR performance due to users being served by a greater number of antennas, when close to a BS.
\bibliographystyle{IEEEtran}
\bibliography{bibliography}

\begin{thebibliography}{10}
\providecommand{\url}[1]{#1}
\csname url@samestyle\endcsname
\providecommand{\newblock}{\relax}
\providecommand{\bibinfo}[2]{#2}
\providecommand{\BIBentrySTDinterwordspacing}{\spaceskip=0pt\relax}
\providecommand{\BIBentryALTinterwordstretchfactor}{4}
\providecommand{\BIBentryALTinterwordspacing}{\spaceskip=\fontdimen2\font plus
\BIBentryALTinterwordstretchfactor\fontdimen3\font minus
  \fontdimen4\font\relax}
\providecommand{\BIBforeignlanguage}[2]{{%
\expandafter\ifx\csname l@#1\endcsname\relax
\typeout{** WARNING: IEEEtran.bst: No hyphenation pattern has been}%
\typeout{** loaded for the language `#1'. Using the pattern for}%
\typeout{** the default language instead.}%
\else
\language=\csname l@#1\endcsname
\fi
#2}}
\providecommand{\BIBdecl}{\relax}
\BIBdecl

\bibitem{LARSSON}
E.~G. Larsson, F.~Tufvesson, O.~Edfors, and T.~L. Marzetta, ``Massive {MIMO}
  for next generation wireless systems,'' \emph{IEEE Communications Magazine},
  vol.~52, no.~2, pp. 186--195, February 2014.

\bibitem{RUSEK}
F.~Rusek, D.~Persson, B.~K. Lau, E.~G. Larsson, T.~L. Marzetta, O.~Edfors, and
  F.~Tufvesson, ``Scaling up {MIMO}: Opportunities and challenges with very
  large arrays,'' \emph{IEEE Signal Processing Magazine}, vol.~30, no.~1, pp.
  40--60, January 2013.

\bibitem{MARZETTA}
T.~L. Marzetta, ``Noncooperative cellular wireless with unlimited numbers of
  base station antennas,'' \emph{IEEE Transactions on Wireless Communications},
  vol.~9, no.~11, pp. 3590--3600, November 2010.

\bibitem{NGO}
H.~Q. Ngo, E.~G. Larsson, and T.~L. Marzetta, ``Energy and spectral efficiency
  of very large multiuser {MIMO} systems,'' \emph{IEEE Transactions on
  Communications}, vol.~61, no.~4, pp. 1436--1149, April 2013.

\bibitem{PITAROKOILIS}
A.~Pitarokoilis, S.~K. Mohammed, and E.~G. Larsson, ``On the optimality of
  single-carrier transmission in large-scale antenna systems,'' \emph{IEEE
  Wireless Communications Letters}, vol.~1, no.~4, pp. 276--279, August 2012.

\bibitem{HOYDIS}
J.~Hoydis, S.~ten Brink, and M.~Debbah, ``Massive {MIMO} in the {UL}/{DL} of
  cellular networks: How many antennas do we need?'' \emph{IEEE Journal on
  Selected Areas in Communications}, vol.~31, no.~2, pp. 160--171, February
  2013.

\bibitem{MULLER}
R.~R. Muller, M.~Vehkapera, and L.~Cottatellucci, ``Blind pilot
  decontamination,'' \emph{International ITG Workshop on Smart Antennas}, March
  2013.

\bibitem{HUH}
H.~Huh, G.~Claire, H.~C. Papadopoulos, and S.~A. Ramprashad, ``Achieving
  "massive {MIMO}" spectral efficiency with a not-so-large number of
  antennas,'' \emph{IEEE Transactions on Wireless Communications}, vol.~11,
  no.~9, pp. 3226--3239, September 2012.

\bibitem{YANG}
H.~Yang and T.~L. Marzetta, ``Performance of conjugate and zero-forcing
  beamforming in large-scale antenna systems,'' \emph{IEEE Journal on Selected
  Areas in Communications}, vol.~31, no.~2, pp. 172--179, February 2013.

\bibitem{OZGUR}
A.~Ozgur, O.~Leveque, and D.~Tse, ``Spatial degrees of freedom of large
  distributed {MIMO} systems and wireless ad hoc networks,'' \emph{IEEE Journal
  on Selected Areas in Communications}, vol.~31, no.~2, pp. 202--214, February
  2013.

\bibitem{JOSE}
J.~Jose, A.~Ashikhmin, T.~L. Marzetta, and S.~Vishwanath, ``Pilot contamination
  and precoding in multi-cell {TDD} systems,'' \emph{IEEE Transactions on
  Wireless Communications}, vol.~10, no.~8, pp. 2640--2651, August 2011.

\bibitem{NGO2}
H.~Q. Ngo and E.~G. Larsson, ``{EVD}-based channel estimations for multicell
  multiuser {MIMO} with very large antenna arrays,'' \emph{Proceedings of the
  IEEE International Conference on Acoustics, Speed and Signal Processing
  (ICASSP)}, pp. 3249--3252, March 2012.

\bibitem{NGO3}
H.~Q. Ngo, E.~G. Larsson, and T.~L. Marzetta, ``Massive {MU-MIMO} downlink
  {TDD} systems with linear precoding and downlink pilots,'' \emph{Allerton
  Conference on Communication, Control, and Computing, Urbana-Champaign,
  Illinois}, pp. 293--298, October 2013.

\bibitem{CHOI}
J.~Choi, Z.~Chance, D.~J. Love, and U.~Madhow, ``Noncoherent trellis coded
  quantization: A practical limited feedback technique for massive {MIMO}
  systems,'' \emph{IEEE Transactions on Communications}, vol.~61, no.~12, pp.
  5016--5029, December 2013.

\bibitem{CHOI2}
J.~Choi, D.~J. Love, and P.~Bidigare, ``Downlink training techniques for {FDD}
  massive {MIMO} systems: Open-loop and closed-loop training with memory,''
  \emph{IEEE Journal of Selected Topics in Signal Processing (J-STSP) on Signal
  Processing for Large-Scale MIMO Communications}, vol.~8, no.~5, pp. 802--814,
  September 2013.

\bibitem{FERNANDES}
F.~Fernandes, A.~Ashikhmin, and T.~L. Marzetta, ``Inter-cell interference in
  noncooperative {TDD} large scale antenna systems,'' \emph{IEEE Journal on
  Selected Areas in Communications}, vol.~31, no.~2, pp. 192--201, February
  2013.

\bibitem{TRUONG}
K.~T. Truong and R.~W. Heath, ``Effects of channel aging in massive {MIMO}
  systems,'' \emph{IEEE/KICS Journal of Communications and Networks, Special
  Issue on Massive MIMO}, vol.~15, no.~4, pp. 338--351, August 2013.

\bibitem{GAO}
X.~Gao, O.~Edfors, F.~Rusek, and F.~Tufvesson, ``Linear pre-coding performance
  in measured very-large {MIMO} channels,'' \emph{Vehicular Technology
  Conference (VTC)}, pp. 1--5, September 2011.

\bibitem{SMITH2}
P.~J. Smith, P.~Dmochowski, M.~Chiani, and A.~Giorgetti, ``On the number of
  independent channels in a diversity system,'' \emph{IEEE Wireless
  Communications and Networking Conference (WCNC)}, pp. 1--6, April 2010.

\bibitem{ZHANG}
J.~Zhang, X.~Yuan, and L.~Ping, ``Hermitian precoding for distributed {MIMO}
  systems with individual channel state information,'' \emph{IEEE Journal on
  Selected Areas in Communications}, vol.~31, no.~2, pp. 241--251, February
  2013.

\bibitem{ZHANG2}
J.~Zhang, Y.~Yu, J.~Mirza, M.~Shafi, M.~Zhang, and P.~A. Dmochowski,
  ``Measurements of {3D} channel impulse response in {C}hina and {N}ew
  {Z}ealand,'' \emph{submitted to IEEE International Conference on
  Communications (ICC)}, June 2015.

\bibitem{MATTHAIOU}
M.~Matthaiou, C.~Zhong, M.~R. McKay, and T.~Ratnarajah, ``Sum rate analysis of
  {ZF} receivers in distributed {MIMO} systems,'' \emph{IEEE Journal on
  Selected Areas in Communications}, vol.~31, no.~2, pp. 180--191, February
  2013.

\bibitem{YIN2}
H.~Yin, D.~Gesbert, and L.~Cottatellucci, ``Dealing with interference in
  distributed large-scale {MIMO} systems: A statistical approach,'' \emph{IEEE
  Journal of Selected Topics in Signal Processing}, vol.~8, no.~5, pp.
  942--953, October 2014.

\bibitem{BASNAYAKA}
D.~A. Basnayaka, P.~J. Smith, and P.~A. Martin, ``Ergodic sum capacity of
  macrodiversity {MIMO} systems in flat {R}ayleigh fading,'' \emph{IEEE
  Transactions on Information Theory}, vol.~59, no.~9, pp. 5257--5270,
  September 2013.

\bibitem{SURAWEERA}
H.~A. Suraweera, P.~J. Smith, and M.~Shafi, ``Capacity limits and performance
  analysis of cognitive radio with imperfect channel knowledge,'' \emph{IEEE
  Transactions on Vehicular Technology}, vol.~59, no.~4, pp. 1811--1822, May
  2010.

\bibitem{SMITH}
P.~J. Smith, C.~T. Neil, M.~Shafi, and P.~A. Dmochowski, ``On the convergence
  of massive {MIMO} systems,'' \emph{IEEE International Conference on
  Communications (ICC)}, pp. 5191--5196, June 2014.

\bibitem{YU}
L.~Yu, W.~Liu, and R.~Langley, ``{SINR} analysis of the subtraction-based smi
  beamformer,'' \emph{IEEE Transactions on Signal Processing}, vol.~58, no.~11,
  pp. 5926--5932, November 2010.

\end{thebibliography}

\appendices
\section{Derivation of Premliminary Result 3}
\label{prelim3_deriv}
Using \eqref{prelim2_start}-\eqref{prelim2_end}, we have
\begin{align}
	\lim_{K\to \infty} \frac{1}{M}\bv^{\textrm{T}}\bP_{i} \bv^{\ast }
	&= \lim_{K\to \infty}\frac{1}{M}\bv^{\textrm{T}}\boldsymbol\phi ^{\textrm{T}}\bQ_{i} \boldsymbol\phi ^{\ast }\bv^{\ast } \\
	&= \lim_{K\to \infty}\frac{1}{M}\tilde{\bv}^{\textrm{T}}\bQ_{i} \tilde{\bv}^{\ast } \\
	&= \lim_{K\to \infty}\frac{1}{M}\sum_{m=1}^{M}{Q_{i,mm}|\tilde{v}_{m}|^{2}} \\
	&= \lim_{K\to \infty}\frac{1}{M}\textrm{tr}(\bQ_{i}) \label{loln} 
\end{align}
\begin{align}
	&= \lim_{K\to \infty}\frac{1}{M}\sum_{n=1}^{N}{\beta _{ni}\textrm{tr}(\boldsymbol\Lambda) } \\
	&= \lim_{K\to \infty} \frac{1}{M}\sum_{n=1}^{N}{\beta _{ni}\left( \frac{M}{N}\right) } \\
	&= \overline{\beta _{i}}, \label{prelim3_end}
\end{align}
where $\tilde{\bv} = \boldsymbol\phi \bv \in \mathbb{C}^{M\times 1}$ has i.i.d. $\mathcal{CN}(0,1)$ elements, $Q_{i,mm}$ is the $(m,m)$th element of $\bQ_{i}$, $\tilde{v}_{m}$ is the $m$th element of $\tilde{\bv}$, and $\overline{\beta _{i}}$ is the average of $\beta _{1i},\beta _{2i},\ldots ,\beta _{Ni}$. Note that \eqref{loln} holds by a version of the law of large numbers for non-identical variables, using $\mathbb{E}\left[ |\tilde{v}_{m}|^{2}\right] =1$.
%
%
\section{Derivation of MF Signal Power}
\label{mf_sig_power}
\begin{align}
	\mathbb{E}\left[ P_{\textrm{sig},i}\right]  
	&= \mathbb{E}\left[ \left| \sqrt{\frac{\rho _{\textrm{f}}}{\gamma }}\bg_{i}^{\textrm{T}}\hat{\bg}_{i}^{\ast }q_{i} \right| ^{2}\right] \\
	&= \frac{\rho _{\textrm{f}}\mathbb{E}\left[ \left| q_{i} \right| ^{2}\right] }{\gamma }\mathbb{E}\left[ \left| (\xi \hat{\bg}_{i}^{\textrm{T}}+\sqrt{1-\xi^{2}}\be_{i}^{\textrm{T}})\hat{\bg}_{i}^{\ast }\right| ^{2}\right] \\
	&= \frac{\rho _{\textrm{f}}}{K\gamma }\left( \xi ^{2} |\hat{\bg}_{i}^{\textrm{T}}\hat{\bg}_{i}^{\ast }|^{2} +(1-\xi ^{2})\be_{i}^{\textrm{T}}\hat{\bg}_{i}^{\ast }\hat{\bg}_{i}^{\textrm{T}}\be_{i}^{\ast }\right) \\
	&= \frac{\rho _{\textrm{f}}}{K \gamma }\left( \xi ^{2} |\hat{\bg}_{i}^{\textrm{T}}\hat{\bg}_{i}^{\ast }|^{2} +(1-\xi ^{2})\textrm{tr}(\hat{\bg}_{i}^{\textrm{T}}\be_{i}^{\ast }\be_{i}^{\textrm{T}}\hat{\bg}_{i}^{\ast })\right) \\
	&= \frac{\rho _{\textrm{f}}}{K \gamma }\left( \xi ^{2} |\hat{\bg}_{i}^{\textrm{T}}\hat{\bg}_{i}^{\ast }|^{2} +(1-\xi ^{2}) \hat{\bg}_{i}^{\textrm{T}}\bP_{i}\hat{\bg}_{i}^{\ast } \right), \label{tmp1a}
\end{align}
where \eqref{tmp1a} is obtained using \eqref{prelim1_end}.
\section{Derivation of MF Interference and Noise Power}
\label{mf_int_power}
\begin{align}
	\mathbb{E}\left[ P_{\textrm{i+n},i}\right] 
	&= \mathbb{E}\left[ \left| \sqrt{\frac{\rho _{\textrm{f}}}{\gamma }}\sum_{k=1,k\neq i}^{K}{\bg_{i}^{\textrm{T}}\hat{\bg}_{k}^{\ast }q_{k}+w_{i}}\right| ^{2}\right] \\
	&= \frac{\rho _{\textrm{f}}}{\gamma }\mathbb{E}\left[ \left( \sum_{k=1,k\neq i}^{K}{\bg_{i}^{\textrm{T}}\hat{\bg}_{k}^{\ast }q_{k}}\right) \left( \sum_{k'=1,k'\neq i}^{K}{\hat{\bg}_{k'}^{\textrm{T}}\bg_{i}^{\ast }q^{\ast }_{k'}}\right) \right] + \sigma ^{2} 
\end{align}
\begin{align}
	&= \frac{\rho _{\textrm{f}}}{K\gamma }\sum_{k=1,k\neq i}^{K}{\mathbb{E}\left[ \left| \bg_{i}^{\textrm{T}}\hat{\bg}_{k}^{\ast }\right| ^{2}\right] } + \sigma ^{2} \\
	&= \frac{\rho _{\textrm{f}}}{K\gamma }\sum_{k=1,k\neq i}^{K}{\mathbb{E}\left[ \left| (\xi \hat{\bg}_{i}^{\textrm{T}}+\sqrt{1-\xi ^{2}}\be_{i}^{\textrm{T}})\hat{\bg}_{k}^{\ast }\right| ^{2}\right] } + \sigma ^{2} \\
	&= \frac{\rho _{\textrm{f}}}{K\gamma }\sum_{k=1,k\neq i}^{K}{\left( \xi ^{2}\left| \hat{\bg}_{i}^{\textrm{T}}\hat{\bg}_{k}^{\ast }\right| ^{2} +(1-\xi ^{2})\mathbb{E}\left[ \hat{\bg}_{k}^{\textrm{T}}\be_{i}^{\ast }\be_{i}^{\textrm{T}}\hat{\bg}_{k}^{\ast }\right] \right) } + \sigma ^{2} \\
	&= \frac{\rho _{\textrm{f}}}{K\gamma }\sum_{k=1,k\neq i}^{K}{\left( \xi ^{2}\left| \hat{\bg}_{i}^{\textrm{T}}\hat{\bg}_{k}^{\ast }\right| ^{2} +(1-\xi ^{2}) \hat{\bg}_{k}^{\textrm{T}}\bP_{i}\hat{\bg}_{k}^{\ast } \right) } + \sigma ^{2}.
\end{align}
\section{}
\label{int_lim_1}
Using $\hat{\bg}_{k}=\bP_{k}^{\frac{1}{2}}\bv_{k}$, $\bP_{k}=\boldsymbol\phi ^{\textrm{T}}\bQ_{k}\boldsymbol\phi ^{\ast }$ and $\tilde{\bv}_{k}=\boldsymbol\phi \bv_{k}$ gives
\begin{align}
	 \lim_{K\to \infty }\frac{1}{M}\left| \hat{\bg}_{i}^{\textrm{T}}\hat{\bg}_{k}^{\ast }\right| ^{2}
	&= \lim_{K\to \infty }\frac{1}{M}\left| \bv_{i}^{\textrm{T}}\boldsymbol\phi ^{\textrm{H}}\bQ_{i}^{\frac{1}{2}}\boldsymbol\phi \boldsymbol\phi ^{\textrm{H}}\bQ_{k}^{\frac{1}{2}}\boldsymbol\phi \bv_{k}^{\ast }\right| ^{2} \\
	&= \lim_{K\to \infty }\frac{1}{M}\left| \tilde{\bv}_{i}^{\textrm{T}}\bQ_{i}^{\frac{1}{2}}\bQ_{k}^{\frac{1}{2}}\tilde{\bv}_{k}^{\ast }\right| ^{2} \\
	&= \lim_{K\to \infty }\frac{1}{M}\tilde{\bv}_{i}^{\textrm{T}}\bQ_{i}^{\frac{1}{2}}\bQ_{k}^{\frac{1}{2}}\tilde{\bv}_{k}^{\ast }\tilde{\bv}_{k}\bQ_{k}^{\frac{1}{2}}\bQ_{i}^{\frac{1}{2}}\tilde{\bv}_{i}^{\ast } \\
	&= \lim_{K\to \infty }\frac{1}{M}\tilde{\bv}_{i}^{\textrm{T}}\bQ_{i}\bQ_{k}\tilde{\bv}_{i}^{\ast } \label{tmp11} \\
	&= \lim_{K\to \infty }\frac{1}{M}\textrm{tr}(\bQ_{i}\bQ_{k}) \label{tmp1} \\
	&= \lim_{K\to \infty }\frac{1}{M}\sum_{n=1}^{N}{\beta _{ni}\beta _{nk}\textrm{tr}(\Lambda ^{2})}.
\end{align}
where \eqref{tmp11} and \eqref{tmp1} hold from the law of large numbers for non-identical variables. Hence,
\begin{align}
	\lim_{K\to \infty }\frac{1}{M^{2}}\sum\limits_{k=1,k\neq i}^{K}{\mathbb{E}\left[ \left| \hat{\bg}_{i}^{\textrm{T}}\hat{\bg}_{k}^{\ast }\right| ^{2}\right] }
	&= \lim_{K\to \infty }\frac{K-1}{\alpha K}\frac{\textrm{tr}(\Lambda ^{2})}{M/N}\sum_{k=1,k\neq i}^{K}{\sum_{n=1}^{N}{\frac{\beta _{ni}\beta _{nk}}{N(K-1)}}} \\
	&= \frac{1}{\alpha }\overline{\Lambda ^{2}}~\overline{\beta _{ik}}.
\end{align}
\section{}
\label{int_lim_2}
Using $\hat{\bg}_{k} = \bP_{k}^{\frac{1}{2}}\bv_{k}$, $\bP_{k}=\boldsymbol\phi ^{\textrm{T}}\bQ_{k}\boldsymbol\phi ^{\ast }$ and $\tilde{\bv}_{k}=\boldsymbol\phi \bv_{k}$ gives
\begin{align}
	\lim_{K\to \infty }\sum_{k=1,k\neq i}^{K}{\frac{\hat{\bg}_{k}^{T}\bP_{i}\hat{\bg}_{k}^{\ast }}{M^{2}}}
	&= \lim_{K\to \infty }\frac{K-1}{\alpha K}\sum_{k=1,k\neq i}^{K}{\frac{\bv_{k}^{\textrm{T}}\bP_{i}\bP_{k}\bv_{k}^{\ast }}{(K-1)M}} 
\end{align}
\begin{align}
	&= \lim_{K\to \infty }\frac{K-1}{\alpha K}\sum_{k=1,k\neq i}^{K}{\frac{ \tilde{\bv}_{k}^{\textrm{T}}\bQ_{i}\bQ_{k}\tilde{\bv}_{k}^{\ast }}{(K-1)M}} \\
	&= \lim_{K\to \infty }\frac{K-1}{\alpha K}\sum_{k=1,k\neq i}^{K}{\frac{\textrm{tr}(\bQ_{i}\bQ_{k})}{(K-1)M}} \label{tmp2} \\
	&= \lim_{K\to \infty }\frac{K-1}{\alpha K}\frac{\sum\limits_{k=1,k\neq i}^{K}{\sum\limits_{n=1}^{N}{\beta_{ni}\beta_{nk}\textrm{tr}(\Lambda ^{2})}}}{(K-1)M} \\
	&= \lim_{K\to \infty }\frac{K-1}{\alpha K}\left( \frac{\textrm{tr}(\Lambda ^{2})}{M/N}\right) \frac{\sum\limits_{k=1,k\neq i}^{K}{\sum\limits_{n=1}^{N}{\beta_{ni}\beta_{nk}}}}{(K-1)N} \\
	&= \frac{1}{\alpha }\overline{\Lambda ^{2}}~\overline{\beta _{ik}}.
\end{align}
Note that \eqref{tmp2} holds by the law of large numbers for non-identical variables and $\overline{\Lambda ^2}$ is the limiting average of the diagonal elements of $\Lambda ^{2}$.

\end{document}